\documentclass{IEEEcsmag}

\usepackage[colorlinks,urlcolor=blue,linkcolor=blue,citecolor=blue]{hyperref}

\usepackage{upmath}

\jvol{XX}
\jnum{XX}
\paper{8}
\jmonth{May/June}
\jname{IT Professional}
\pubyear{2019}

\definecolor{cb_orange}{rgb}{1.0,0.51,0.0}
\definecolor{cb_blue}{rgb}{0.22,0.49,0.72}
\definecolor{cb_green}{rgb}{0.3,0.67,0.29}
\definecolor{cb_red}{rgb}{0.89,0.1,0.11}
\definecolor{cb_purple}{rgb}{0.6, 0.31, 0.64}
\definecolor{cb_brown}{rgb}{0.6, 0.4, 0.2}
\definecolor{cb_crimson}{rgb}{0.86, 0.08, 0.24}

\begin{document}

\sptitle{Department: People in Practice}
\editor{Editors: Melanie Tory, m.tory@northeastern.edu and Dan Keefe, dfk@umn.edu}

\title{The Ball is in Our Court: Conducting Visualization Research with Sports Experts}

\author{Tica Lin}
\affil{Harvard University}

\author{Zhutian Chen}
\affil{Harvard University}

\author{Johanna Beyer}
\affil{Harvard University}

\author{Yingcai Wu}
\affil{Zhejiang University}

\author{Hanspeter Pfister}
\affil{Harvard University}

\author{Yalong Yang}
\affil{Virginia Tech}

\markboth{People in Practice}{Visualization Research in Sports}

\begin{abstract}
Most sports visualizations rely on a combination of spatial, highly temporal, and user-centric data, 
making sports a challenging target for visualization. 
Emerging technologies, such as augmented and mixed reality (AR/XR), have brought exciting opportunities along with new challenges for sports visualization. 
%As sports data are spatial, temporal, and user-centric, emerging technologies such as augmented and mixed reality has brought exciting opportunities and new challenges for visualization researchers. 
We share our experience working with sports domain experts and present lessons learned from conducting visualization research in SportsXR. 
In our previous work, we have targeted different types of users in sports, including athletes, game analysts, and fans. 
Each user group has unique design constraints and requirements, 
such as obtaining real-time visual feedback in training, 
automating the low-level video analysis workflow, 
or personalizing embedded visualizations for live game data analysis.
In this paper, 
we synthesize our best practices and pitfalls we identified while working on SportsXR. 
We highlight lessons learned in working with sports domain experts in designing and evaluating sports visualizations and in working with emerging AR/XR technologies.
We envision that sports visualization research will benefit the larger visualization community 
through its unique challenges and opportunities for immersive and situated analytics.

%Through synthesizing our pitfalls and best practices, we highlight lessons learned in working with sports domain, design and evaluation of sports visualization design study, and working on emerging technologies. We envision further development in sports visualization research can bring novel aspects to visualization research community through unique visualization challenges.

% Based on a user-centered design methodology, we developed visualization systems to augment different sports experts' workflows. 
% For each target user and task, we tailored research methods to adapt to research goals and practical constraints
% in requirement identification, visualization design, and evaluation.
% Through synthesizing our pitfalls and best practices, 
% we highlight six key research objectives and challenges across all study phases and exemplify various considerations from our previous work. 
% Finally, we suggest ten guidelines to highlight the distinction in conducting user-centered design studies in the sports domain.    
\end{abstract}

\maketitle

\begin{figure*}[t!]
    \centering
    \includegraphics[width=.85\textwidth]{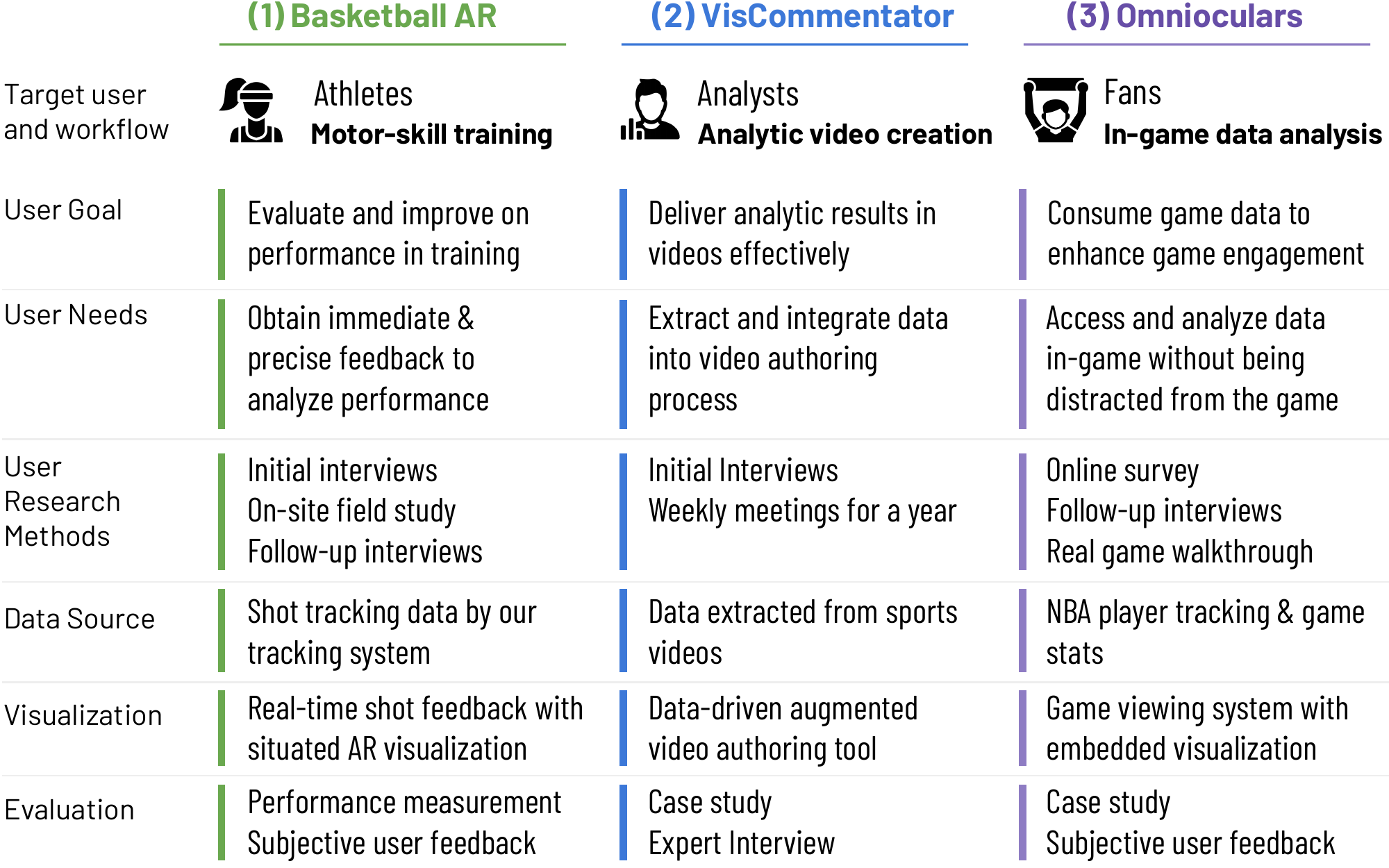}
    \caption{Three exemplary sports visualization projects we discuss in Sec 2. Each project targets a different user group and workflow, including athletes' motor-skill training, analysts' video creation, and fans' in-game data analysis. Different user goals and needs lead to tailored research and design approaches.} 
    \label{fig:projects}
    \vspace{-4mm}
\end{figure*}

\chapterinitial{Data} analytics in sports can be a game changer, 
as it enables its stakeholders to make clear-sighted decisions based on objective evidence~\cite{perin2018state}.
One of the most successful stories is the MLB Moneyball, where Oakland Athletics General Manager Billy Beane, with a limited budget, used sabermetrics (i.e., a specific statistical analysis method in baseball) to measure player performance. He then used discovered insights to draft undervalued players and subsequently won games against rich teams like the New York Yankees. 
In modern sports, data-driven decision-making has become ubiquitous in most processes, such as player recruitment, performance evaluation, training, and tactic analysis~\cite{baca2009ubiquitous}. 

However, data-driven decision making and sports data analysis is still often difficult to access for sports professionals.
Imagine a college basketball player struggling to make free throws. 
The coaches tell them to improve their shooting trajectories, 
but that is hard to achieve when athletes are mostly practicing on their own.
How could athletes get immediate and data-driven feedback directly on the court?
Moreover, 
consider a basketball fan watching a game starring their favorite player. 
It is the 2nd quarter and the player is not performing well. 
The fan is getting worried and pulls up their phone to check the player's recent performance, but it is hard to find. 
Being distracted, the fan misses an all-star play of their favorite athlete, easily the best moment of the game. 
How could this game-watching experience have benefited from easily accessible data-based insights presented situated right next to the player?
Lastly, consider a data analyst working for a basketball team. 
After collecting and analyzing spatio-temporal data for days, they have found some exciting insights. 
However, when presenting statistics and abstract graphs, the coaches have a hard time understanding the point and dismiss the insights. 
How could this data have been presented closer to its spatial and temporal context, to make insights easier to understand?

What do these three scenarios have in common? 
First, they are all based on real stories from sports practitioners we have worked with. 
Second, the visualizations in these three scenarios
% More importantly and fortunately, modern data-driven and situated visual analytics approaches bring the possibilities to solve all of them.
% Data visualization in sports 
all come with 
a set of unique constraints 
in \emph{space}, \emph{time}, and \emph{user}
that make their adoption difficult.
% \emph{Spatial} constraints, where the sports data originates from a physical space, and should be displayed in that same physical context to aid understanding. 
% \emph{Temporal} constraints, in scenarios where people need to access data in a timely manner. 
% And \emph{user} constraints, where diverse groups of users in sports have different needs for data and visualizations.
These constraints might not be present in all sports applications, however, when they are, conventional visualization solutions may no longer work.
%Not all sports applications have these constraints, and not all constraints apply to any single sports application. 
%However, when these constraints are present in a sports application, conventional visualization solutions may no longer work.
This is mainly because conventional sports analytics is often performed \emph{off-line} and \emph{off the court}.
As a result, users cannot access the visualization at their preferred location and time, forcing users to change their workflow or making some application scenarios impossible.
% The key unique \emph{user} challenge in sports visualization is that different user groups have little overlap in their requirements.
% In many other domains, the requirements of different user groups form a containment relationship, where one user group needs either more or less features than another group.
% Sports visualization applications, on the other hand, require designers and developers to very clearly identify the differences in user needs from different groups, to sufficiently support their workflows.

On the positive side, sports visualization is currently reaching a pivotal moment in time.
Novel sensors and data collection techniques and recent progress in human-computer interaction and mixed reality offer unprecedented possibilities for modern sports visualizations.
For example, 
state-of-the-art computer vision (CV) technologies provide \emph{real-time} data collection \emph{in the court}, 
allowing applications that would not be possible without access to such data;
% Natural language processing and machine learning models enable users to express their intent in more intuitive ways and to even predict user intentions, to reduce the number of interactions required;
the next generation of display and interaction devices (i.e., virtual and augmented reality or VR/AR) are capable of projecting an unlimited number of interactive 2D and 3D graphics into any space, and breaking the boundary between the physical and digital world~\cite{lin2020sportsxr}.
With these advances in technology, 
basketball players who want to improve their shooting can obtain objective feedback through CV and AR techniques in real-time~\cite{lin_towards_2021};
% no longer need to subjectively \emph{feel} their trajectories, 
% instead, we can use CV to track the ball, and present the 3D trajectories for their point of view in AR in real-time 
fans who want to access external information during a play can see the data embedded in the court~\cite{lin2023quest};
% voice their questions 
% and the relevant data and visualization will then be embedded in the court
and analysts who want to better explain their insights can create situated sports videos for more effective communication~\cite{chen_augmenting_2022}. 

The purpose of this paper is to discuss the opportunities and challenges in modern sports visualization (or SportsXR~\cite{lin2020sportsxr}) research.
We first elaborate the spatial, temporal, and user constraints to highlight the uniqueness of modern sports visualizations.
We then share three projects we performed with different sports user groups (see Fig.~\ref{fig:projects}) and finally discuss the lessons learned and potential future challenges we see from these projects.

\section{WHAT MAKES SPORTS DIFFERENT?}
\label{sec:constraints}

%Sports have unique traits that lead to distinct considerations when designing visualizations for sports domain users.
%When designing visualizations for sports data and sports domain users, there are several unique considerations 
Sports data, as well as sports domain users, have some unique traits that need to be considered when designing sports visualizations.

\noindent\textbf{Spatial constraints.} 
Most sports data are by nature \emph{spatial}, either containing spatial coordinates, such as the position of the basketball and players, or by being linked to physical objects or spaces~\cite{willett2016embedded}.
For example, shooting percentages of a basketball player are linked to specific areas on the basketball court.
Moreover, sports data are typically dynamic~\cite{yao2022visualization}, with most entities (e.g., the players) moving all the time, making it challenging to design visualizations that allow viewers to easily track these changes. 

\noindent\textbf{Temporal constraints.}
Most sports are fast-paced, creating constant streams of time-sensitive data, such as updated player positions or velocities.
Sports visualizations have to be able to support these time-sensitive data, to allow viewers to follow the temporal aspect of the sport. This requires visualizations to run at real-time on high-throughput data streams. At the same time, however, sports visualizations must limit the mental load on viewers, to allow them to digest the presented information in a limited time.
Additionally, sports visualizations should not distract users from their primary tasks, which might be coaching a team, watching a game on TV, or practicing free shots. Instead of competing for their attention, sports visualization needs to be integrated into the user's workflow to support real-time data analysis.

\noindent\textbf{User constraints.} 
%There are multiple user groups in the sports. The more c
Common stakeholders in sports include athletes, coaches, analysts, fans, and journalists, and less common roles can be managers, sports doctors, and more. 
The diverse requirements of different user groups lead to completely different sports visualization designs.
For \textsc{athletes}, visualizations should enable them to connect the data back to their physical skills~\cite{lin_towards_2021} (e.g., to improve their technique or tactics).
For \textsc{coaches}, visualizations should be deployed on mobile devices, to be looked at while being close to their athletes, next to the court, rather than in front of an office computer.
\textsc{Analysts} consider both the exploratory and explanatory aspects of sports data and need to be able to quickly and easily create different visualizations based on their specific purposes~\cite{chen_augmenting_2022}.
%Meanwhile, creating these visualizations is usually challenging for them.
The wide variety of \textsc{Fans}, from novices to die-hard fans, adds even more challenges to the design of visualizations~\cite{lin2023quest}.
Finally, \textsc{journalists}
leverage visualizations to perform in-depth sports data exploration and communicate the generated data story to their readers~\cite{fu_supporting_2022}.
These user groups vary widely in their visualization literacy and analytical background, but also in their motivation and the data they are interested in.
%in, and their visualization goals. 
Therefore, it is essential that visualization tools are customized to specific user groups, their data, and the goals they are trying to achieve.

\noindent\textbf{Summary.} 
% From an overview of the constraints, 
The spatial and temporal constraints are often interconnected, and modern sports visualization has the potential to address them together.
For example, creating visualizations situated in physical space allows viewers to consume spatially-embedded sports data faster and with less cognitive load. % than using another digital device (e.g., a mobile phone).
Other domains, outside of sport, share similar challenges and could benefit from SportsXR research, such as visualizing spatio-temporal data.
We also foresee that applying existing techniques from computer graphics and computer vision, like optimizing camera placement and rendering, can advance sports visualization in certain scenarios.
Some challenges have bigger implications but have been underexplored, such as the limited amount of time for users to perceive visual information, which has only been discussed preliminarily~\cite{lin2020sportsxr,yao2022visualization}. 
%Other topics, such as the temporal constraints we discussed, have broader semantics that are underexplored, such as the limited amount of time for users to perceive visual information, which has only been discussed preliminarily~\cite{lin2020sportsxr,yao2022visualization}.

% \input{body/2-background}
\section{SPORTSXR RESEARCH}
In the following, we summarize three sports visualization projects that we have conducted with athletes, analysts, and fans, respectively (see Fig.~\ref{fig:projects}).
% We followed a user-centered design process
% ~\cite{abras2004user} 
% based on the design study methodology. 
For each project, we summarize our initial user research, identified goals \& tasks, design considerations, and evaluation.
%
%
%%jj 
%The first project focuses on athletes practicing free throws with the main goal of providing real-time feedback on shot performance directly on the court, to improve motor-skill learning.
%jj 
%The second project focuses on enabling analysts to create augmented sports videos to support a data-driven workflow, rather than a completely manual annotation process.
% The second project focuses on improving the workflow of analysts creating augmented sports videos 
% with the main goal of supporting a data-driven workflow, rather than a completely manual annotation of videos.
% The third project focuses on sports fans and how to improve their live game watching experience. Here, the main goal was to understand and categorize user needs, and to design a prototype for embedded visualizations in live game watching setups.
%jj
%The third project focuses on improving sports fans' live game watching experience by rendering embedded visualizations.
% Here, the main goal was to understand and categorize user needs, and to design a prototype for embedded visualizations in live game watching setups.

\subsection{Basketball AR: Motor-skill training for athletes}
% goal & audience
The goal of this project was to provide real-time visual feedback
%In this project, we designed real-time visual feedback
% to support situated analytics 
for basketball players practicing free-throw shooting. 
Through close collaboration with college basketball teams, we prioritized a minimal visual design and fast and accurate feedback to help improve players' motor skills~\cite{lin_towards_2021}.

% % how did you identify the requirement
\textbf{Interviewing athletes to identify the gap in shot training.} 
Through user interviews and in-person field studies with players and coaches, 
% we closely observed the training workflow of athletes on the basketball team 
% to identify potential needs for visualization research. 
% During the study,
% We started out by meeting with coaches to ask about their daily practices, 
% focusing on activities that either involved data analytics or that could be enhanced by data visualization. 
% Followed by in-person site visits to observe the team practice, 
we found players desire precise quantitative feedback on their shots, which they typically do not get. % but do not have access to them.
% we identified a gap in free-throw training as players desire precise quantitative feedback on their shots in real time but do not have access to them.
% Most players practice shooting on their own 
% and only subjectively evaluated their shot performance by mentally tracking the trajectory of their shots
% and made adjustments on their shooting form based on \textit{``their feeling''}. 
They also have no way to precisely specify goals or evaluate their outcome in an actionable way, i.e., players cannot easily evaluate the shot angle of their free throws. 
Therefore, players were highly interested in accessing visual feedback during shot training.

% design requirement
% Based on the study, %in-depth field study and interviews, 
We first characterized user goals and tasks to form a set of visualization design requirements. 
Basketball players wanted to have real-time quantitative feedback to increase shot consistency and refine their shot arc.
% The goals of basketball players were to get real-time quantitative feedback, increase shot consistency, and refine their shot arc. 
We then identified four tasks a visualization system for free-throw training needs to support: 
T1) analyzing the target shooting arc before a shot, 
T2) analyzing one's shooting arc during and after each throw immediately, 
T3) comparing one's shooting arc to the target arc, 
and T4) adjusting one's target arc to get consistently closer to an ideal arc.

\begin{figure}[t!]
    \centering
    \includegraphics[width=\columnwidth]{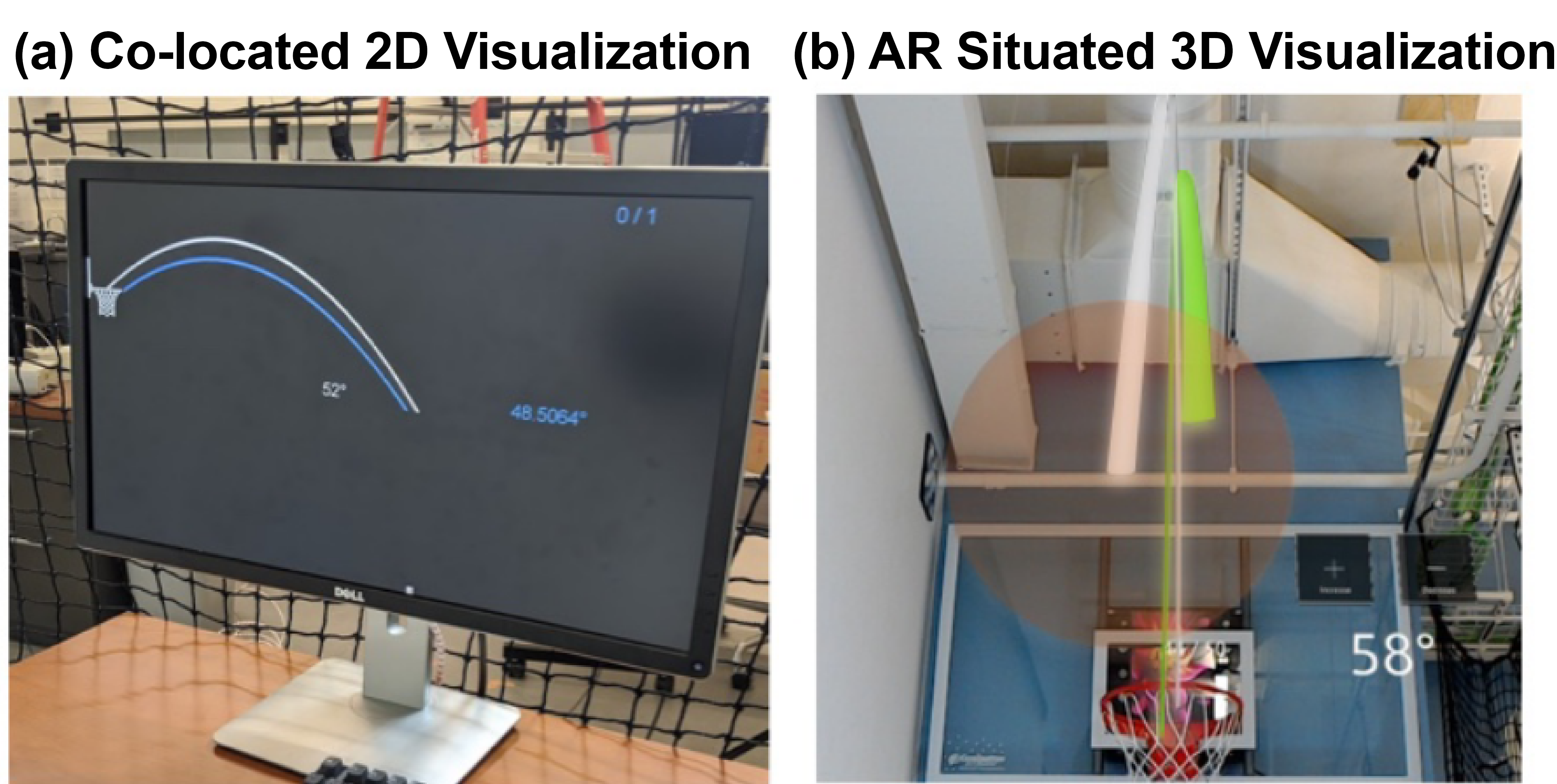}
    \caption{2D and AR free-throw shot visualizations.}
    \label{fig:arshot}
    \vspace{-4mm}
\end{figure}

% design description
\textbf{Real-time visual feedback through co-located and situated visualization.}
To fulfill the aforementioned design requirements, the visualization needs to support the three-dimensional reasoning of the shot, be located closely and easily accessible to the user.
% , and provide immediate visual feedback.
% 
We implemented a shot tracking system to detect shot metrics in real-time
and designed 
% To design our shot arc visualization,
% we iterated our design ideas to address identified design challenges and ended up with 
two visualization modalities. 
A \textit{co-located 2D visualization} (Fig.~\ref{fig:arshot}a) presents the shot arc from a third-person view on a co-located monitor placed next to the shooter. 
% The player adjusts the target arc by changing the launch angle using a keyboard. 
A \textit{situated AR 3D visualization} (Fig.~\ref{fig:arshot}b) presents the shot arc from a first-person view through an AR headset. 
% The player adjusts the target shot angle and release point using a keyboard in 2D and gestures in AR interface.
Both visualizations present information about the shooter's target and actual shot arcs with 
different presentation dimensions (2D vs. 3D) and viewing perspectives (third-person vs. first-person).

% To fulfill the aforementioned design requirements, our visualization needed to support spatial reasoning of the shot, be located closely and easily accessible to the user, and provide immediate visual feedback.

% We first implemented a shot tracking system to detect shot metrics in real-time.
% For visual feedback we designed a co-located 2D visualization (Fig.~\ref{fig:arshot}a) and a situated 3D visualization in AR (Fig.~\ref{fig:arshot}b).
% The latter shows the shot arc in-situ in a first-person view through a AR headset. The player can adjust the target shot angle and release point through a gesture-based AR interface.
% Both visualizations present information about the shooter's target and actual shot arcs with different interaction and presentation dimensions.

% evaluation
\textbf{Comparing the usefulness of 2D and AR visualizations for free-throw training.}
%As AR visualization was considered a novel technique, 
Our evaluation focused on characterizing unique aspects of applying our novel AR visualization to free-throw training. 
% To this end, 
We conducted a comparative study of free-throw shot training with the 2D and AR visualizations with ten basketball players over four days each. 
We collected both quantitative and qualitative measures, such as shot metrics and user feedback.
% , including shot metrics, feedback throughout the study and follow-up survey. Through the mixed method data analysis, we capture users' shot improvement and insights.
Our results suggest that the 2D and AR visualizations were considered useful for free-throw training, and users significantly improved their shot angle consistency throughout the study.
More interestingly, the AR visualization promotes an increased focus on body form as users prioritized ``improve the shooting form'' as their top training goal, as opposed to ``improve shot accuracy'' with the 2D visualization.

\subsection{VisCommentator: Augmented sports video creation for analysts}
\begin{figure}
    \centering
    \includegraphics[width=\columnwidth]{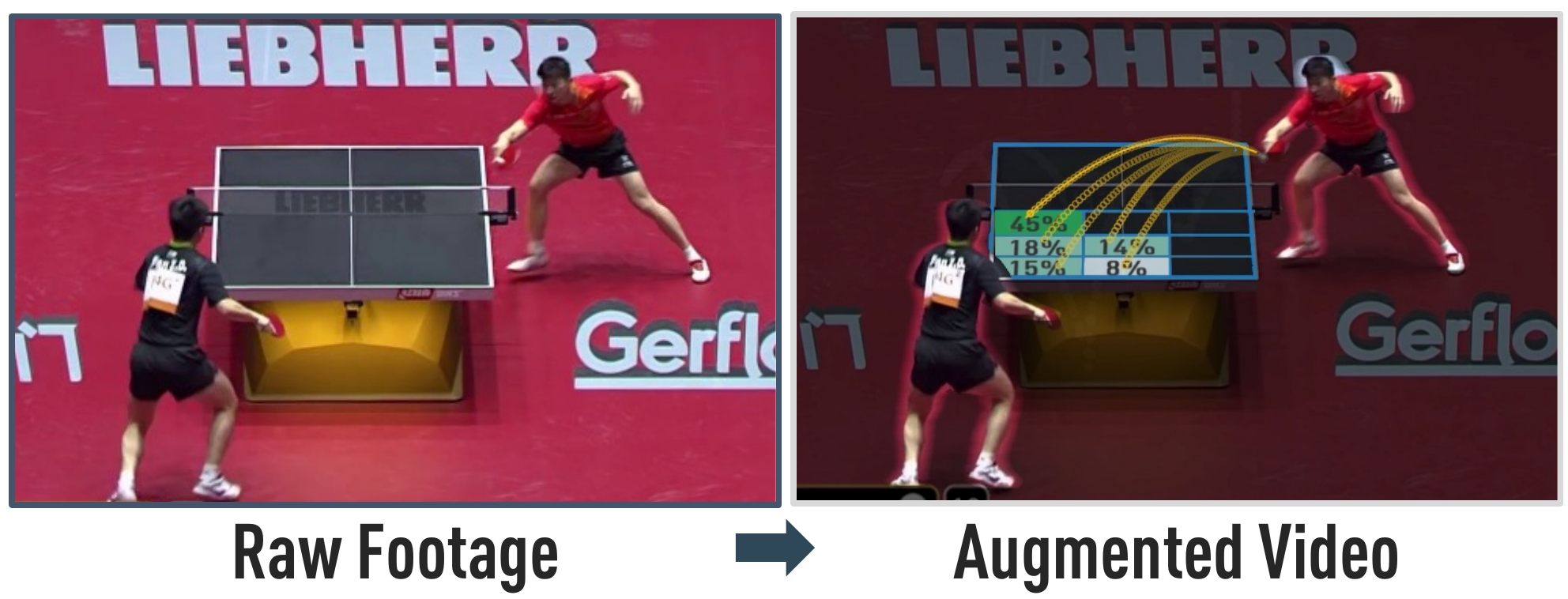}
    \caption{Augmented sports videos embed visualizations into videos to present data.}
    \label{fig:arvideo}
    \vspace{-4mm}
\end{figure}

In this project, 
we focused on enabling analysts to create augmented sports videos in a data-driven workflow (see Fig.~\ref{fig:arvideo}) rather than a completely manual annotation process~\cite{chen_augmenting_2022}.
%In this project, we developed a tool for sports analysts to better communicate their insights by creating sports videos with embedded visualizations~\cite{chen_augmenting_2022}.
% In our second study, we focused on sports analysts and developed a tool for
% creating augmented sports videos, which embed visualizations into videos, to communicate the analysts' insights~\cite{chen_augmenting_2022}.
%. We collaborated with two domain experts over the course of 11 months to

% how did you identify the requirement
\textbf{A Formative Study to Understand Constituents of Augmented Sports Videos.}
Augmented sports videos help in communicating analytical insights. 
% \yy{One more sentence to elaborate the reason. E.g., as people can digest information more effectively with sufficient spatial-temporal contexts.} 
However, creating augmented sports videos is often difficult,
involving complex design decisions and video editing.
To understand the design practices of augmented sports videos,
we collected and analyzed 233 videos from TV, teams, and leagues.
Based on our analysis, 
we propose a design space that characterizes augmented sports videos at element- and clip-levels with four design dimensions (i.e., Data Type, Visual Type, Data Level, and Narrative Order).
% In weekly meetings with two experts over the course of eleven months, 
% we iterated on our design goals and demonstrated several prototype systems.

\begin{figure*}[t!]
    \centering
    \includegraphics[width=\textwidth]{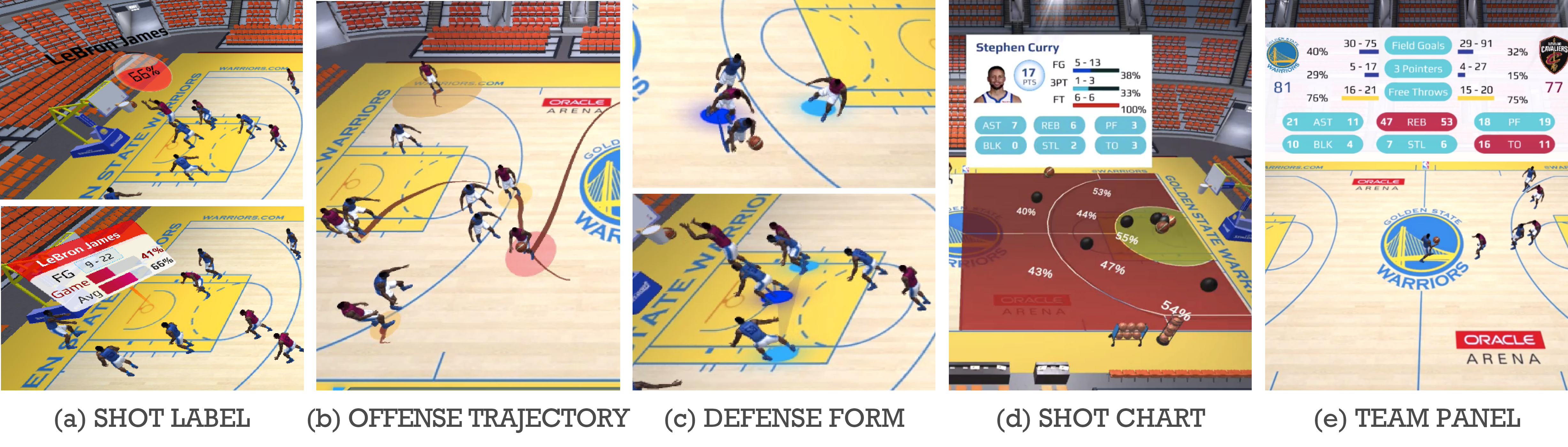}
    \caption{Embedded visualizations for five selected game contexts, including (a) shooting performance, (b) offense player trajectories, (c) defenders' movement, (d) player performance, and (e) team performance.}
    \label{fig:omni}
    \vspace{-4mm}
\end{figure*}

% design requirement
\textbf{A Semi-Automated System to Create Augmented Sports Videos.}
Informed by the design space and the close collaboration with experts,
we developed \emph{VisCommentator},
a system that allows sports analysts to
augment table tennis videos efficiently by selecting the data to visualize instead of manually drawing graphical marks.
Given a raw table tennis video,
VisCommentator first extracts sports data (e.g., players and ball positions, ball events) using a set of machine learning models.
The extracted data are grouped together based on their semantic levels.
Next, VisCommentator uses the extracted data to decorate the video objects
and the video timeline,
% attach the extracted data to the video objects
so that users can directly interact with the video objects and events
to select the data to visualize.
% select the data to visualize
% by directly interacting with the video objects. 
Finally, 
VisCommentator suggests visual effects for the user-selected data based on our design space
and renders the visual effects into the video.
% based on the narrative order and data selected by the use.
% we map the data to visuals by using a recommendation system developed based on our design space.

% and user preferences.
% we identified three design goals for the fast prototyping tool,
% including \emph{G1. extract the data from the video based on data levels},
% \emph{G2. interact with the data instead of graphical marks},
% and \emph{G3. recommend visualizations for different narrative purposes}.
% design description
% \textbf{Design Considerations and Outcomes.}
% Based on the design goals, 

% evaluation
\textbf{Expert evaluation.}
To evaluate the usability of \emph{VisCommentator},
we conducted a qualitative user study with seven sports analysts.
% The study aimed to evaluate whether sports analysts can create augmented table tennis videos with our system and to observe their creation process to reflect on future improvements.
We tasked participants with reproducing target augmented videos and recorded the completion time and success rate for each task, 
and subjective feedback.
All participants could successfully reproduce the augmented videos. 
\subsection{Omnioculars: In-game data analysis for fans}

In this project, we focused on improving the experience of sports fans in live game watching scenarios by embedding visualizations into basketball games.

% goal
%Our third study targets visualizations for general sports audiences during basketball game watching. 
%To enhance fans' data understanding and engagement,
We investigated the design space of embedding visualizations into live game videos and developed an interactive game-viewing prototype with personalized embedded visualizations to support fans' in-game data analysis~\cite{lin2023quest}.

% With diverse fan groups and game situations, we emphasized the personalized and context-dependent aspects of embedded visualizations for in-game data analysis.

% how did you identify the requirement
\textbf{Characterizing basketball fans' data needs during a live game.}
% To understand how fans inquire data throughout live game watching, 
We first deployed an online survey 
to collect fans' data needs in a live game
and interviewed six fans about their game viewing experiences.
% including their data-seeking moments, type of data, and tasks they seek to accomplish. 
% We then followed up with six fans for in-depth interviews to understand their current workflows, pain points, and best practices of looking up data in games.
% interview insights details
% Throughout our formative study, 
We found that fans seek data in various contexts during games, which we characterized 
% We further characterized the contexts of fans' data needs 
into three components, 
including \textsc{Scenarios} (\textit{when} fans desire to look for data), %, comprising a combination of a target (player or team), event and time), 
\textsc{Data} (\textit{what} in-game data fans look up), % i.e., box scores, tracking data, and video), 
and \textsc{Tasks} (\textit{why} fans look for data). % i.e., identify, compare, and summarize the performances of teams and players). 
Based on this framework, we derived design requirements under specific game contexts.
All interviewees confirmed their habits of looking up game stats on separate screens during live games, which introduces context switching and reduces game engagement.
% to complement their game viewing experiences. 
% mainly because data provided in the live streaming videos lack diversity and controllability. 
%However, looking up data on a separate screen suffer from context switching, which reduces game engagement. 

\textbf{Design exploration of embedded visualizations in sports games.}
%To fill the gap in consuming game data in a live game, 
To present contextual data with embedded visualization in game videos, we designed an interactive game viewing system, \textit{Omnioculars}.
Based on the design feedback from fans, we prioritized five game contexts to drive our embedded visualization designs,
% To prioritize the game contexts to be designed for, we extracted 20 contexts from the formative study records and sketched design mockups to gather fans' feedback. 
% Based on fans' ratings and comments, five primary contexts were selected to drive our embedded visualization designs, 
including shooting, offense, defense, player performance, and team performance
% We then developed \textit{Omnioculars} to present embedded visualization designs for the five selected contexts in a game simulator, 
(Fig.~\ref{fig:omni}).

% evaluation
\textbf{Using embedded visualizations for in-game data analysis.}
To evaluate our system for in-game data analysis, we conducted a user study with simulated basketball game clips in two parts. The participants evaluated each of the five embedded visualizations and 
% to derive game insights and evaluate its usefulness and engagement. 
then freely combined different visualizations to derive game insights. 
% Participants rated the system's usefulness and reasoned about their strategies and preferences of using different embedded visualizations.

Participants considered all five embedded visualizations useful and novel. 
% Each embedded visualization provides helpful data on distinct game aspects according to our design contexts. 
Moreover, participants used different interaction approaches  based on their game focus and preferences (i.e., configuring a fixed set of visualizations ahead of time or altering visualizations on-the-fly based on context). They could derive distinct game insights with their chosen embedded visualizations. Overall, they found Omnioculars helpful and fun to use, felt in control of their experience, and were likely to use it in live games.

\section{LESSONS LEARNED}

% We summarize the lessons learned from our projects into two themes.

% Based on our experiences,
% we found three particular difficulties in working on visualization research for sports, 
% including the high entry barrier to the sports domain, the difficulties to propose changes to alter the sports industrial standard, and the highly personalized user needs.

% -------------- Working with sports experts
% Required domain knowledge
\subsection{Theme 1: Working with Sports Experts}

%to start working in sports domain
\textbf{The entry barrier is high}
due to the \emph{required domain knowledge}, 
\emph{seasonality of sports leagues},
and \emph{limited data availability}.

% Without our track record in sports, it could be challenging to convince the sport's practitioners to engage with us.
Without personal connections and a deep understanding of the sport, 
it is difficult to find collaborators and keep them engaged. 
For our basketball AR project, 
the collaboration with the varsity teams only started because the lead author had experience working with a professional basketball team and was introduced to the coach through a mutual connection.
Similarly, 
the augmented video project was only initialized because of the researcher's prior experience with table tennis visual analytics systems.
%since one of the coauthors has already developed many visual analytics systems for table tennis.
At the very least,
coaches, analysts, or fans usually expect the researchers to have a good understanding about the sport, 
such as the rules of the game and the composition of the professional league or collegiate tournament. 
For example, 
when interviewing NBA fans, 
they often refer to specific players or teams and technical terms (e.g. pick-and-roll, floater) to describe their game-viewing experiences. 
It is crucial that researchers show understanding and enthusiasm to encourage the conversation and extract more insights from the interviewees. 

In addition, sports leagues are seasonal and follow strict timelines.
% such as an upcoming tournament or a game season.
Researchers need to be aware of the constrained availability of experts when planning project phases.
For example, in the basketball AR project, 
the varsity teams were able to support us throughout the design phases, 
but they were competing in a tournament at the later phase. 
% Therefore, coaches introduced us to other intramural teams to support our user testing. 
Therefore, we had to find other intramural teams to support our user testing. 
% Our Omnioculars project was conducted when the NBA season just started.
% in October
We planned our Omnioculars project to align with the NBA season to increase our chances to find passionate fans and keep them in the loop throughout the entire project. 
% This helped us find passionate fans and keep them in the loop throughout the entire project. 

Lastly, access to sports data is usually proprietary to the team or league. 
Since the nature of sports is competitive, teams and athletes try to get advantages over others through data collection and analysis. Therefore, sharing data 
% with researchers for academic purposes 
is usually not encouraged or even impossible. 
For example, all teams we collaborated with explicitly asked us not to publicize their data and process.
% while varsity teams we worked with were willing to share data they collected to help us understand their current workflow, they explicitly asked us not to publicize their data and process due to privacy and competitiveness reasons. 
Similarly, when designing embedded visualizations for basketball game viewing, we could not obtain up-to-date player tracking data from the NBA and had to use previously collected game data.
% and official broadcasting companies
% To address such a gap in data, we either need to build our own camera tracking system to gather motion data or train computer vision models to extract data from the game videos.
A workaround could be to build our own camera tracking system and train computer vision models to extract data. However, this would require expertise in CV and considerable resources.

% adoption varies
\textbf{Educating sports experts to use new techniques is hard}
since some experts are \emph{reluctant to try alternatives} 
while others may \emph{overestimate the capabilities of emerging techniques}.

% \textbf{Sports experts have a well-established workflow and rarely accept alternatives.}
% and it is challenging to suggest alternatives.}
% Unlike other professionals, 
% \zt{one sentence here}
Oftentimes, sports experts have well-established workflows and are reluctant to consider alternatives.
% investigate the incentives of using new technologies. 
% often are reluctant to try alternatives enabled by new technologies.  
% for their whole life and thus often 
% Many professional coaches were athletes prior to becoming coaches and may stay in the same domain for most of their life.
% For these coaches, using new technologies usually means they will have to learn new things, step out of their comfort zone, and take the risk of gaining no benefits. 
% This sometimes makes it difficult to convince them to change their current workflow.
%Taken together with the high entry barrier to sports domain, 
%this means it could be difficult to propose changes to their current workflow.
In VisCommentator, even though our proposed authoring system can support fast prototyping of augmented videos,
some analysts still preferred using traditional court diagrams.
% to communicate tactics and findings with the players.
% \zt{one sentence of solution: what should we do? be patient, empathetic, and persuasive?}
To this end, we sought to find motivating factors in the target users. 
For example, collegiate teams usually are open to collaborating with school research labs because innovation appeals to student athletes, 
which may also enhance their recruitment process.
% because of the long history of the sports industries' standard.
% \tl{For example, even though tablet-based tactic boards were designed to support faster iteration~\cite{Seidl2018}, most professional basketball coaches still use a traditional whiteboard to draw tactics and communicate them with the players.}

On the other hand,
sports experts may overestimate
% and have impractical expectations from emerging technologies.
% of 
the capabilities of emerging technologies that are still in an evolving stage but are advertised as omnipotent (e.g., universal artificial intelligence). %This leads sports experts to unrealistic expectations of what the current technologies can achieve.
In the AR free-shot project, for example, 
some experts were disappointed as they found that the HoloLens1 was far from mature and thus provided biased negative feedback for the visualizations in the user study. %VisCommentator
It is crucial to properly \emph{manage experts' expectations} of the emerging techniques
% , especially when they are not the main focus of the novel visualization system 
to obtain unbiased and useful feedback.

\subsection{Theme 2: Evaluating Immersive Visualizations}
% Most of our collaboration ended up becoming a design study of visualization research. 
% Below, we shared challenges we encountered and methods we applied in designing.

% \textbf{Identifying the visualization contribution and limiting the scope of the implementation are important.}
% %Due to the data availability problem mentioned in the previous section, 
% Many design problems are hindered by having no or limited data. In that case, advanced computer vision and machine learning techniques, 
% such as tracking players in the game videos, are needed to extract the necessary data.
% Additionally, 
% unlike desktop environments,
% which most visualization systems are developed for,
% immersive environments (e.g., AR, VR) are still in their infancy
% and lack accessible visualization toolkits.
% Taking these issues together, developing immersive sports visualizations is particularly challenging.
% However, in visualization research, 
% we are interested in human factors like how to design visualization to increase data understanding and decision making of domain experts. 
% Therefore, we found it necessary to 
% emphasize the visualization research contribution
% and properly scope the system implementation.

\textbf{Identifying the benefits of visualizations in the analytic workflow.} A key consideration during the design process is to distinguish between the added values from the data versus visualization designs. When presented with a novel analytic system, sports experts may find the system useful because of the data they were able to obtain. 
To identify the design values, it is important to evaluate them explicitly. 
In our free-shot project, we present the same shot tracking data in the 2D and AR visualizations. The direct comparison between different visual representations allowed us to separate the benefit of immersive visualization, leading to more generalizable design guidelines.
% for other sports analytic applications.

% You can only do “one” thing at a time in VR/AR
% VR/AR may not have the other software you commonly use on a desktop/or they are not optimized for VR/AR (e.g., through a mirrored virtual screen)
% Users have to learn a lot about the device and the specific application we provide

\textbf{Using Wizard-of-Oz user testing allows us to focus on the visualization research questions.}
% can ease the research of immersive sports visualizations.} 
% to evaluate complex sports visualization system
Considering the high implementation costs of AR,
we advocate for developing a subset of target scenarios and using the Wizard-of-Oz method for evaluation. 
% complex sports visualization system.
In the VisCommentator project, 
% we implemented a thorough data processing pipeline and demonstrated the feasibility of our proposed design 
we focused on a selected game video instead of trying to implement a complete solution for all game videos.
On the other hand, in the Omnioculars project, we aimed to evaluate how people use embedded visualizations to analyze a basketball game. 
% However, without accurate player tracking and camera configuration data, it was technically challenging to embed visualizations into real game videos. 
Instead of tackling CV problems beyond our expertise, we designed a simulated game environment with 3D models and manually crafted player animation for selected game scenarios. We also used a Wizard-of-Oz method to evaluate how users interact with different visualizations through verbal commands. 
In both cases, we have obtained valuable insights from users interacting with our visualization tools under realistic use cases without being blocked by implementation details.

\textbf{Evaluating sports visualizations requires consideration for individual user differences and contexts.} 
% Due to the diverse user needs in sports, 
% the same visualization tool might not be useful for all users on the same task, e.g., training free-throw shooting with situated AR visual feedback or analyzing games with embedded visualizations. 
% Furthermore, as many factors could impact the sports performance, such as a different opponent or game strategy, 
% it is difficult to properly evaluate the usefulness of visualizations using quantitative measurement with simplified tasks. 
% \tl{
%With the diverse user needs and varying factors that could impact sports performance, i
It is often insufficient to evaluate sports visualizations using general quantitative measurement with simplified tasks.
% }
One mitigation is to use qualitative methods across user groups, contexts, and times. 
For example, we evaluated Omnioculars with novice and hard-core fans and recorded their interaction patterns. 
We also evaluated visualizations under different game contexts, such as under shooting or clutch time scenarios.
In AR basketball training, we collected user feedback at the beginning, during, and after training. This allowed us to evaluate the usefulness of visualizations at a finer granularity with considerations for various factors on top of the standard quantitative measures. 

% \zt{Limited empirical knowledge/evidence in VR/AR visualizations}
% Graphical Perception for Immersive Analytics: https://cmci.colorado.edu/visualab/3DPerception/3DPerception.pdf

\textbf{Evaluating skill transfer and long-term adoption are still lacking in sports visualizations.} 
As much as we are excited about the new SportsXR solutions, it is difficult to convince sports experts to change their workflow without evidence for long-term improvement and adoption. 
Long-term evaluations of visual analysis tools for sports are still lacking.
%Evaluating sports visual analysis tools for long term is still lacking. 
%jj This can be due to the lack of incentives for the sports experts and the uncertain impact of the tool on athletes' performance,
%as sports experts seek mature technologies to ensure their success. 
However, to have a meaningful impact on the sports domain, it is necessary to show evidence of skill transfer through longitudinal evaluation. 
We thus envision and advocate for sports visualization research to expand its impact through collaborations with
other research areas, such as Kinesiology and Sports Sciences.

\section{CONCLUSIONS}

% In this paper, we provided practical guidance on conducting visualization research with sports experts. We drew upon our previous experience in working directly with various target users in the sports domain. 
% \tl{We identified lessons learned in working with sports domain experts, designing and evaluating visualizations, and working on emerging technologies. By sharing the unique challenges faced by visualization researchers working in the sports domain, we present our best practices and pitfalls, and hope to inspire impactful future sports visualization research.}

Exploring new possibilities in sports visualization with modern computing and display technologies is extremely exciting, especially when seeing its impact on sports practitioners. %the WOW moments from sports practitioners. 
However, designing SportsXR applications is not a trivial task. 
%and comes with unpredicted challenges. 
%jj
%In this paper, we have discussed the best practices and pitfalls of conducting sports visualization research and have drawn upon our previous experience working with various types of sports experts. 
Sports is a data-rich scenario for visualization that comes with many specific design considerations and constraints.
% when designing sports visualizations.
Although there are some inspiring discussions and preliminary studies ~\cite{willett2016embedded,yao2022visualization,ens2021grand}, fundamentally, it is still unclear how existing visualization principles can best be applied to modern sports visualizations, and there is a lack of established guidelines.
We need to invest more in understanding human factors in sports visualizations and closely collaborate with other communities (e.g., CV, ML, NLP) to improve the current workflows in sports.
% and even create novel and better ones.
% More research is needed in...to...
Furthermore, we envision that with visualization becoming ubiquitous in space and time, there will be more and more scenarios that share similar challenges as we have discussed here for sports. 
Ultimately, research in different applications can complement empirical knowledge, guidelines, and techniques from different perspectives, forming a complete ecosystem for future visualization applications.

% In this paper, we discussed best practices and pitfalls of conducting visualization research with sports experts. We drew upon our previous experience in working directly with various types of practitioners in the sports domain. 

% \tl{We identified lessons learned in working with sports domain experts, designing and evaluating visualizations, and working on emerging technologies. By sharing the unique challenges faced by visualization researchers working in the sports domain, we present our best practices and pitfalls, and hope to inspire impactful future sports visualization research.}

% characterized six key research objectives in three primary phases of a user-centered design studies---
% identifying requirements, designing, and evaluating visualizations. 
% By 
% examining pitfalls and best practices for each objective, we present ten guidelines for tailoring and carrying out research methods  
% to ensure the success of collaborating with sports domain experts and designing novel visualizations.

% What can sports vis bring to vis:
%  * Inform immersive/situated vis and interaction design
%  * Whether previous empirical results apply to our new problems? New human factors to investigate.
%  * Inform new evaluation metrics (no longer just time and accuracy)

\section{ACKNOWLEDGMENT}

This work was partially supported by NSF grants III-2107328 and IIS-1901030.

\bibliographystyle{unsrt}
\bibliography{references}

\begin{IEEEbiography}{Tica Lin}{\,}is a Ph.D. student at the Visual Computing Group at Harvard University. 
% Prior to joining Harvard, she was a UX Developer at NBA 76ers. Her research interests include data visualization, immersive analytics, and human-computer interaction. In particular, she explores novel visualization and interaction design in Mixed and Augmented Reality.
Contact her at mlin@g.harvard.edu.
\end{IEEEbiography}
\vspace{-2mm}

\begin{IEEEbiography}{Zhutian Chen}{\,} is a postdoc at the Visual Computing Group at Harvard University. 
% Before joining Harvard, he was a postdoc at University of California San Diego, and a Ph.D. student at Hong Kong University of Science and Technology. His interests are in Information Visualization, Human-Computer Interaction, and Augmented Reality. 
Contact him at ztchen@seas.harvard.edu.
\end{IEEEbiography}
\vspace{-2mm}

\begin{IEEEbiography}{Johanna Beyer}{\,}is a research scientist at the Visual Computing Group at Harvard University. 
% Before
% joining Harvard, she was a postdoctoral fellow at the
% Visual Computing Center at KAUST. She received
% her Ph.D. in computer science at the University of
% Technology Vienna, Austria, in 2009. Her research
% interests include immersive analytics, scalable methods for visual abstractions, and large-scale volume visualization.
Contact her at jbeyer@g.harvard.edu.
\end{IEEEbiography}
\vspace{-2mm}

\begin{IEEEbiography}{Yingcai Wu}{\,} is a Professor at the State Key Lab of CAD\&CG, Zhejiang University. 
% His main research interests are in information visualization and visual analytics, with focuses on sports science and urban computing. He received his Ph.D. degree in Computer Science from the Hong Kong University of Science and Technology. Prior to his current position, Dr. Wu was a postdoctoral researcher at the University of California, Davis from 2010 to 2012, and a researcher in Microsoft Research Asia from 2012 to 2015. 
Contact him at ycwu@zju.edu.cn.
\end{IEEEbiography}
\vspace{-2mm}

\begin{IEEEbiography}{Hanspeter Pfister}{\,}is the Academic Dean of Computational Sciences and Engineering and An Wang Professor of Computer Science at Harvard University. 
% He has a Ph.D. in Computer Science from Stony Brook University and an MS degree in electrical engineering from ETH Zurich, Switzerland. His research in visual computing lies at the intersection of visualization, computer graphics, and computer vision and spans a wide range of topics, including biomedical visualization, image and video analysis, machine learning, and data science.
Contact him at pfister@g.harvard.edu.
\end{IEEEbiography}
\vspace{-2mm}

\begin{IEEEbiography}{Yalong Yang}{\,}is an Assistant Professor in the Department of Computer Science at Virginia Tech. 
% He
% was a Postdoctoral Fellow at the Visual Computing
% Group at Harvard University, and a Ph.D. student
% at Monash University, Australia. In his research, he
% designs and evaluates interactive visualizations on
% both conventional 2D screens and in 3D immersive
% environments (VR/AR). 
Contact him at yalongyang@vt.edu.
\end{IEEEbiography}

\end{document}